\documentstyle[12pt,aasms4]{article}

\lefthead{M. Reyes-Ruiz}

\righthead{Magnetorotational instability in the dead zone}

\begin{document}

\title{The magnetorotational instability across the dead zone of 
 protoplanetary disks}

\author{M. Reyes-Ruiz}
\affil{Instituto de Astronom\'{\i}a, UNAM, Apdo. Postal 877,
Ensenada, B.C. 22800, M\'{e}xico.}

\begin{abstract} 
We examine the linear stability of a flow threaded by a weak,
vertical magnetic field in a disk with a keplerian rotation profile and
a vertical stratification of the ionization degree as that predicted 
for vast portions of protoplanetary disks. A quasi-global analysis is
carried out, where the form of the perturbations in the vertical
direction is determined. Considering the ohmic magnetic diffusivity 
of the gas, the conditions leading to the magnetorotational
instability are analyzed as a function of the diffusivity at
the disk surfaces, its vertical profile and the strength of the
unperturbed magnetic field. For typical conditions believed to prevail in
protoplanetary disks at radial distances between 0.1 and 10 AU, where the
so-called dead zone is proposed to exist, we find that generally the
instability is damped. This implies that, if the
MRI is considered the only possible source of turbulence in
protoplanetary disks, no viscous angular momentum transport occurs
at those radii.  
\end{abstract}

\keywords{accretion, accretion disks -- magnetic fields -- 
MHD -- planetary systems -- solar system: formation}

Disks around classical T-Tauri stars, also called protoplanetary 
disks (or PP disks hereafter), are commonly
modeled as viscous accretion disks. 
Observed evolutionary trends for 
the global properties of PP disks have been shown to be 
roughly consistent with those predicted by accretion disk models 
(Hartmann et al. 1998, Stepinski 1998). However, the acceptance 
of such models is justifiably not 
universal and their precise properties are far from established.
Nevertheless, partially on account of the existing theoretical 
framework, accretion disk models constitute the basis of most 
current models for protoplanetary disks. 

In such models, 
molecular viscosity being drastically 
insufficient to explain the observed evolutionary timescales for 
these objects, an anomalous viscosity is invoked to drive 
the evolution of PP disks. Albeit their precise origin was not readily
identified, turbulence and magnetic fields where suggested, and generally
accepted, as responsible of the required anomalous ``viscous'' torques
(Lynden-Bell 1969, Shakura \& Sunyaev 1973, Eardley \& Lightman 1976). 
In the seminal work of Balbus \& Hawley
(1991) the instability of perfectly conducting, keplerian flows, 
threaded by a weak magnetic field, was identified as an unavoidable 
factor in the dynamics and magnetic field generation process in 
accretion disks. Since that time, this so-called
magnetorotational instability (also called MRI hereafter) has been 
shown to lead to
self-sustained MHD turbulence capable of transporting angular momentum 
outwards through the disk while mass is driven toward the central
object (Hawley, Gammie \& Balbus 1995,  Brandenburg et al. 1995, 
Stone et al. 1996). 

As long as magnetic diffusive processes are small enough and a weak magnetic
field threads a keplerian disk, the MRI arises and leads to
self-sustained turbulence capable of driving disk evolution 
(Jin 1996, Balbus \& Hawley 1998). These conditions are generally met in 
accretion disks around compact objects but apparently not in extended 
regions of protoplanetary disks (Gammie 1996, D'alessio et al. 1998). 
Temperatures less than 1000 K outwards of $\sim 1$ AU predicted by models 
of PP disks result in a sharp drop of the ionization degree (Stepinski 1992).
Since the magnetic ohmic diffusivity is inversely
proportional to the ionization degree this implies the importance of
diffusive processes beyond such distances (see also Stepinski, Reyes-Ruiz
\& Vanhala 1993). There, galactic cosmic rays and other external 
sources become the principal ionizing agents. 

An important feature of
the ionization due to such extraneous factors is that, since their 
effect is shielded as they travel into the
disk, the ionization fraction is strongly stratified decreasing 
from the disk surfaces to the midplane (Dolginov \& Stepinski 1994). 

Under the assumption that the MRI is the only source of self-sustained
turbulence and ``viscous'' torques in accretion disks, 
Gammie (1996) proposed that mass transfer in PP disks, beyond 
the inner hot region, takes place in 
a layered manner. In the layered accretion scenario angular momentum 
and mass transfer occurs only through active layers near the 
disk surfaces, where the magnetic diffusivity is low enough for 
the MRI to develop, while the gas stagnates in the strongly 
diffusive ``dead zone'' around the midplane. The evolution of 
protoplanetary disks via layered accretion has been calculated by 
Stepinski (1999) finding that it can be dramatically 
different from that predicted by ``standard'' models
of PP disks, with vertically uniform viscosity, as those 
presented by Ruden \& Lin (1986), 
Ruden \& Pollack (1991) and Reyes-Ruiz \& Stepinski (1995) 
among others.
 
Given the importance of having reliable, detailed models 
for the structure and evolution of PP disks as foundation for
theories of planet formation, we believe a critical revision 
of the layered accretion scenario is justified. We begin 
this work in the present paper by studying the effect of 
diffusivity stratification on the linear development of the MRI.

We consider the most important implicit assumptions 
leading to the scenario of layered accretion 
(Gammie 1996) are the following. 
First, that the MRI develops in the active regions leading to 
MHD turbulence, and angular momentum transport, as it does in 
homogeneous disks, i.e. disks without diffusivity stratification. 
Secondly, the stagnant condition giving 
the dead zone its name in the model of
Gammie (1996) follows from the assumption that no significant 
stresses, of Reynolds or Maxwell type, are present in such
region. The first assumption is motivated by the 
results of the local, linear analysis carried out by Jin
(1996) who found that although ohmic magnetic diffusivity 
can quench the MRI, at least for the minimum mass solar 
nebula model, the ionization degree in the active
layers may be sufficient for the instability to develop.
However, as discussed by Sano \& Miyama (1999) a quasi-global 
analysis, incorporating the boundary
conditions of the problem in the $z$ direction and
the stratification of the magnetic diffusivity, could yield 
new constraints on the instability criterion. Moreover, at
least this level of nonlocality in the analysis of the MRI
is required if one is to say anything about the velocity and
magnetic field perturbations, and resulting stresses, 
induced in the dead zone of PP disks.

The aim of this paper is to analyze in detail 
the emergence of the MRI across the dead zone of
protoplanetary disks. We address the main assumptions of
the layered accretion scenario outlined above. In 
contrast to the recent, similar work by Sano \& 
Miyama (1999) we adopt the simplest geometry and 
perturbation type for which the MRI is known to arise
and focus 
on the specific effects of a magnetic diffusivity
profile as that expected in standard, accretion disk 
models of PP disks. We seek to analyze the 
stability criterion and 
$z$ dependence of the perturbations through the active 
layers and dead zone. The implications 
of our results on the dynamics of PP disks
are discussed for a wide range of disk 
models. 

\section[]{Formulation of the problem}

We consider the quasi-global, linear stability of a 
keplerian flow threaded by a uniform vertical magnetic 
field in a medium with $z$-dependent ohmic diffusivity. 
Our quasi-global study follows in spirit the analysis 
by Gammie \& Balbus (1994) but we include a variable 
ohmic diffusivity in the magnetic induction equation. 
This configuration is considered as a model of the middle 
portions of protoplanetary disks where the dead zone is 
proposed to exist (Gammie 1996). As a first approximation 
to this problem, for reasons purely of mathematical
simplicity, we consider only the effect of ohmic magnetic 
diffusion. The influence of other diffusive processes
on the MRI, such as ambipolar diffusion, 
has been studied by Wardle (1997, 1999) in a local approach. 
Although such processes are probably important in 
regions of PP disks, their incorporation into a  
global analysis is left for future contributions. 
 
In our quasi-global analysis the unperturbed system 
is considered homogeneous in the radial and azimuthal 
directions in all properties except the angular velocity,
which is a function of the radial distance from the 
central star. We concentrate on the effect of 
diffusion on the most unstable modes found in the 
ideal MHD analysis, those corresponding to destabilized 
Alfven waves traveling in the $z$-direction 
(Balbus \& Hawley 1998). To this end we assume 
an unperturbed weak magnetic field configuration 
like ${\mathbf B} = B \hat{z}$ where $B$ is a constant. 
The equilibrium condition for the flow, in the presence 
of the gravitational force due to the central star, 
corresponds to circular rotation with velocity 
${\mathbf U} = r \Omega \hat{\phi}$, where $\Omega$ is 
equal to $\sqrt{G M_\star}/r^{3/2}$. This is the
simplest disk configuration where the instability 
arises in the ideal MHD analysis (Balbus \& Hawley 1998).     

Our starting point are the governing continuity equation, 

\begin{equation}  
\frac{\partial \rho}{\partial t} \ + \
{\mathbf \nabla} \cdot (\rho {\mathbf U}) 
\ = \ 0 ,   
\label{conti}  
\end{equation}

\noindent momentum conservation equation, 

\begin{equation}  
\frac{\partial {\mathbf U}}{\partial t} \ + 
({\mathbf U} \cdot {\mathbf \nabla} {\mathbf U}) \ =
\ - \frac{1}{\rho} {\mathbf \nabla} \left( P + 
\frac{|{\mathbf B}|^2}{8 \pi} \right) 
\ + \ \frac{1}{4 \pi \rho} ({\mathbf B} \cdot 
       {\mathbf \nabla} {\mathbf B})
\ - \ {\mathbf \nabla} \Phi ,   
\label{momen}  
\end{equation}

\noindent where $\Phi$ is the gravitational potential from the 
central star, and magnetic induction equation,

\begin{equation}  
\frac{\partial {\mathbf B}}{\partial t} \ = \
 {\mathbf \nabla} \times ({\mathbf U} \times {\mathbf B} 
\ - \ \eta  {\mathbf \nabla} \times {\mathbf B}),   
\label{induc}  
\end{equation}

\noindent where $\eta$ is the ohmic magnetic diffusivity, a
function of $z$ in general. The system physical 
properties; velocity, magnetic 
field, density, $\rho$, and pressure, $P$, are perturbed 
with axisymmetric linear disturbances of
the form  $ g(z) {\rm e}^{-\gamma t}$, where $g(z)$
represents the $z$-dependent amplitude of the 
perturbation on any of the physical properties
of the system. 

In the approximation for geometrically thin disks
and weak magnetic fields, the restrictions mentioned above  
allow us to reduce the governing equations, 
(\ref{conti})-(\ref{induc}),
to a system of coupled equations similar to (8)-(14) of 
Sano \& Miyama (1999). An important simplification 
in our approach is that, in absence of a 
toroidal component of the background field, $B_{\phi}$, 
the $r$ and $\phi$ components of the momentum and
magnetic induction equations, containing the Alfven 
modes, decouple from the density, pressure and 
vertical velocity perturbations. 

We represent the perturbations of velocity and 
magnetic field components as $u_r, \ u_\phi$ 
and  $b_r, \ b_\phi$ respectively. There is no radial
dependence of the perturbation since we restrict our 
analysis to vertically traveling perturbations. 
Future contributions will deal with a more general 
derivation.  
 
To first
order, we can write the equations for 
the amplitudes of perturbed quantities as:

\begin{equation}  
- \gamma u_r \ - \ 2 \Omega u_{\phi} \ - \ \frac{B}{4\pi \rho} D b_r  
\ = \ 0 ,  
\label{ur8}  
\end{equation}  
  
\begin{equation}  
- \gamma u_{\phi} \ + \frac{\Omega}{2} u_r \ - \ \frac{B}{4\pi  
\rho} D b_{\phi} \ = \ 0  , 
\label{uphi8}  
\end{equation}  
  
\begin{equation}  
- \gamma b_r \ - B D u_r \ - [ \eta D^2 + D\eta D ] b_r \ = \ 0  , 
\label{br8}  
\end{equation}  
  
\begin{equation}  
- \gamma b_{\phi} \ - \ B D u_{\phi} \ + \ \frac{3}{2} \Omega b_r  
- [ \eta D^2 + D\eta D ] b_{\phi} \ = \ 0  ,  
\label{bphi8}  
\end{equation}  

\noindent where $D$ represent the operator $d/dz$. 
The density, $\rho$ appearing in equations 
(\ref{ur8})-(\ref{bphi8}) is the
unperturbed value, taken from the condition of 
hydrostatic equilibrium in the $z$ direction for an
isothermal disk,
  
\begin{equation}  
\rho (z) \ = \ \rho_o {\rm e}^{-z^2/H^2} ,    
\label{rho1}  
\end{equation}

\noindent where $\rho_o$ is the value of the density at the
disk midplane and $H = \sqrt{2} C_s/\Omega$ is the isothermal
scale height, with $C_s$ the sound speed in the gas.

These system of equations is solved subject to the following 
boundary conditions:
 
\begin{equation} 
\left. D u_r \right|_{z=\pm H_t} = \left. D u_\phi \right|_{z=\pm H_t} = 0 
,
\label{bcsu} 
\end{equation} 
  
\begin{equation} 
\left. b_r \right|_{z=\pm H_t} = \left. b_\phi \right|_{z=\pm H_t} = 0 
,
\label{bcsb} 
\end{equation} 

\noindent where $H_t$ is the height above the midplane where 
the boundary conditions are applied. These boundary conditions 
correspond to a hot, tenuous halo model for the exterior of the disk,
in a force-free state
with the assumption of vanishing stress as $|z| \rightarrow \infty$.
Such model has been previously used
in similar quasi-global analysis of the MRI (Gammie \& Balbus 1994,
Sano \& Miyama 1999). The derivation,  
presented by Gammie \& Balbus (1994), is simplified 
to equations (\ref{bcsu}) and (\ref{bcsb}) in our case 
of perturbations dependent only on $z$. We will return to discuss
the assumed boundary conditions in section 4.

\section{Numerical solution}
 
To ease manipulation in numerically computing the growth rates of the 
instability, $\gamma$, 
we first normalize all variables by dividing by $\Omega$ and making
the
following substitutions:

\begin{equation}
\hat{\gamma} = \gamma / \Omega,
\end{equation}
\begin{equation}
\hat{b}_i = b_i / B \ \ {\rm for} \ i = r, \phi,
\end{equation}
\begin{equation}
\eta = \eta_o f(z),
\end{equation}
\begin{equation}
\hat{z} = z / H ,
\end{equation}
\begin{equation}
\hat{u}_i = u_i / C_s \ \ {\rm for} \ i = r, \phi,
\end{equation}
\begin{equation}
\beta =  C_s^2 / V_A^2 \ \ {\rm where} \ V_A^2 = B^2/4\pi\rho,
\end{equation}
\noindent and
\begin{equation}
R_m = H^2 \Omega / \eta_o 
.
\end{equation}
 
Our parameter $\beta$ is that typically used in
plasma physics and our $R_m$ is a magnetic Reynolds 
number based on the velocity $C_s$ over a scale $H$. 
Note that this differs slightly from the traditional definition 
of $R_m$ in terms of the flow velocity appearing in the magnetic
induction equation. Since $\rho$
depends on $z$ in our model, so does $\beta$. We write 
$\beta = \beta_c / h(z)$, where $\beta_c$ is the value of the
parameter at the disk midplane and $h(z) = {\rm e}^{\hat{z}^2}$ 
in our assumed isothermal disk structure.
The
resulting system of equations can be written as
 
\begin{equation} 
- \hat{\gamma} \hat{u}_r \ - \ 2 \hat{u}_{\phi} \ - \ 
  \frac{1}{\sqrt{2}} \beta_c^{-1} h D \hat{b}_r  \ = \ 0 ,  \label{ur9} 
\end{equation} 
 
\begin{equation} 
- \hat{\gamma} \hat{u}_{\phi} \ + \frac{1}{2} \hat{u}_r \ - \ 
   \frac{1}{\sqrt{2}} \beta_c^{-1} h D \hat{b}_{\phi} \ = \ 0  , 
\label{uphi9} 
\end{equation} 
 
\begin{equation} 
- \hat{\gamma} \hat{b}_r \ -  \frac{1}{\sqrt{2}} D \hat{u}_r \ - 
  R_m^{-1} [ f D^2 + Df D ] \hat{b}_r \ = \ 0   ,
\label{br9} 
\end{equation} 
 
\begin{equation} 
- \hat{\gamma} \hat{b}_{\phi} \ - \  \frac{1}{\sqrt{2}} D \hat{u}_{\phi} \ 
  + \ \frac{3}{2} \hat{b}_r  - R_m^{-1} [ f D^2 + Df D ] 
\hat{b}_{\phi} \ = \ 0.  
\label{bphi9} 
\end{equation} 

The convenience of this normalization is making explicit the dependence 
of the solution to our problem on the parameters $\beta_c$ and $R_m$, it
will also depend on the form of the diffusivity profile and indirectly 
on the boundary conditions.

Now we discretize the variables and coefficients on a set of points
$\hat{z}[i] \in [-\hat{H_t},\hat{H_t}]$ 
using the notation $g[i]$ to denote the value of
the variable $g$ at point $i$. The derivatives $D$ and $D^2$ of the
variables are 
approximated with 2nd order centered finite differences on an
equispaced grid ($\hat{z}[i] = \hat{z}_{\rm min} + i * \Delta \hat{z}$). 
With this, the system of 4 differential equations (\ref{ur9})-(\ref{bphi9}) 
becomes a system of 4$N$ coupled algebraic equations, where $N$ is the number
points in the grid. Corresponding to each point there are 4 equations so that
the discretized system of difference equations can be recast in matrix 
form as 
 
\[ 
{\mathbf A} 
 \left( \begin{array}{c}
 \hat{u}_r[1]  \\
 \hat{u}_\phi[1] \\
 \hat{b}_r[1] \\
 \hat{b}_\phi[1] \\
 \cdot \\  \cdot \\  \cdot \\
 \hat{u}_r[N]  \\
 \hat{u}_\phi[N] \\
 \hat{b}_r[N] \\
 \hat{b}_\phi[N]
\end{array} \right) 
\ - \ \hat{\gamma}
 \left( \begin{array}{c}
 \hat{u}_r[1]  \\
 \hat{u}_\phi[1] \\
 \hat{b}_r[1] \\
 \hat{b}_\phi[1] \\
 \cdot \\  \cdot \\  \cdot \\
 \hat{u}_r[N]  \\
 \hat{u}_\phi[N] \\
 \hat{b}_r[N] \\
 \hat{b}_\phi[N]
\end{array} \right) 
 \ = \ 0 
\]

\noindent where the growth rate of our perturbations, $\hat{\gamma}$, are
seen to be the eigenvalues of the matrix $\mathbf A$ made up of the
corresponding terms from the discretization of the 
system of equations (\ref{ur9})-(\ref{bphi9}). 

We calculate eigenvalues numerically transforming the matrix to upper
Hessenberg form and using a QR algorithm. The number of points in the
grid was varied from 10 to 50 with no significant change (less than 1\%)
in the value
of the maximum growth rate of the instability. Results shown correspond
to a discretization over 20 points for $\hat{z}$.

The boundary conditions are
incorporated rewriting the corresponding 
equations (\ref{ur9})-(\ref{bphi9}) for the first and last points 
of our grid,
upon discretization of the conditions (\ref{bcsu}) and (\ref{bcsb}) 
with 2nd order forward and backward differences at the bottom and top 
disk surfaces respectively.

\section{Results}

With the application to the middle, poorly ionized 
portions of PP disks in mind,
we compute growth rates and eigenmodes (the
form of the perturbations) for weak unperturbed magnetic fields ($\beta_c
\geq 0.1$) and moderately diffusive disks.  
We consider the later condition as $ 1 < R_m \leq 10^{3}$  
since the parameter $R_m$ 
can be seen as the ratio of the characteristic diffusive timescale 
over a length $H$ to the dynamical timescale.

\subsection{Case of uniform diffusivity}

A global analysis for a homogeneous disk with ohmic diffusivity
has recently been presented by 
Sano \& Miyama (1999). Here we repeat their calculation following a
slightly different approach, and a different computational method, 
to ease the comparison to the results for a stratified disk.

\begin{figure}  
\plotone{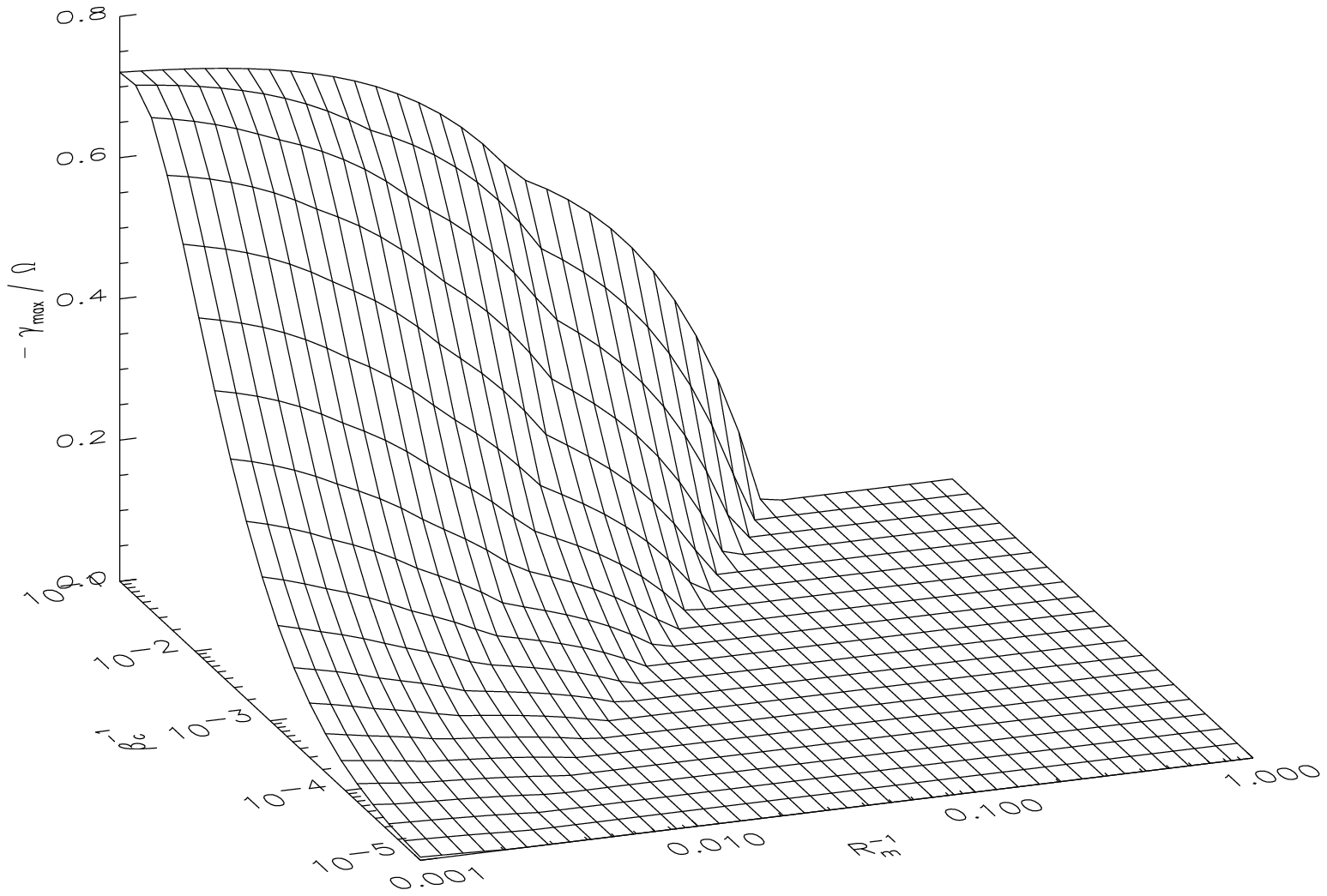}  
\caption{Maximum growth rate of the magnetorotational 
instability as a function of 
$\beta_c$ and $R_m$, in a medium with uniform ohmic
diffusivity.}   
\label{fig1}  
\end{figure}

To consider a homogeneous diffusivity we simply put $Df = 0$
and $f = 1$ in the system (\ref{ur9})-(\ref{bphi9}).	
Figure \ref{fig1} shows the value of the maximum growth rate of the
instability, given by the minimum value of $\hat{\gamma}$ with no
imaginary part, as a
function of the resistivity reflected in
$R_m$, and the
unperturbed magnetic field strength, entering through 
the parameter $\beta_c$.

Fitting a line through the nodes of the function depicted in Figure 
\ref{fig1} indicates that the MRI arises as long as the following 
relation holds:

\begin{equation}
\beta_c < R_m^2 
\label{coneta1}
\end{equation}

\noindent with the stated weak field, weak diffusion restrictions,

\begin{equation}
\beta_c > 1 
\ {\rm and} \ R_m 
> 1
\end{equation}
 
We can recast condition (\ref{coneta1}) in terms of the 
characteristic timescales
of the problem: the diffusion timescale, 
$t_{\rm dif} = H^2/\eta$ and the
Alfven wave $H$-crossing time, $t_{\rm A} = H/V_A$. For a density
stratified disk with constant magnetic diffusivity, the MRI will
arise at a given radius if:

\begin{equation}
t_{\rm A} \lesssim t_{\rm dif}
\label{coneta2}
\end{equation}

In order for the instability to develop an Alfven wave, with the
velocity at the disk midplane, must have 
time to cross the disk before the field diffuses over the same
length scale. This result, summarized in  
condition (\ref{coneta1}), is equivalent to the requirement 
that the Lundquist number be greater than unity. The 
Lundquist number is the ratio of the Alfven velocity to the 
``velocity'' of drift of magnetic field lines through the 
conductor due to diffusion.
These results are consistent with Sano \& Miyama (1999). 

\subsection{Effect of diffusivity stratification}

The calculation of growth rates for a stratified disk is now a simple
procedure of prescribing a particular analytic profile for $f(z)$ and 
following the steps described above. 

As we have mentioned we expect
the diffusivity to increase rapidly from its minimum value at the disk 
surfaces. The results we present consider a $z$-dependence of
the ohmic diffusivity like:

\begin{equation} 
f(\hat{z}) = {\rm e}^{(\hat{H}_t^2-\hat{z}^2)} \ 
       {\rm e}^{({\rm Erf}(\hat{H}_t) - {\rm Erf}(\hat{z}))/l_o}
\label{perfil}
\end{equation} 
 
\noindent where Erf refers to the well known Error function. 
The precise form
for this profile and particular 
choices for $l_o$ will be justified in the discussion section.
In addition to $\beta_c$ and $R_m$,
in a disk with diffusivity stratification two new parameters 
$l_o$ and $H_t$, determine
the diffusivity profile at a given radius in the disk and
the growth rates of the MRI.

\begin{figure}  
\plotone{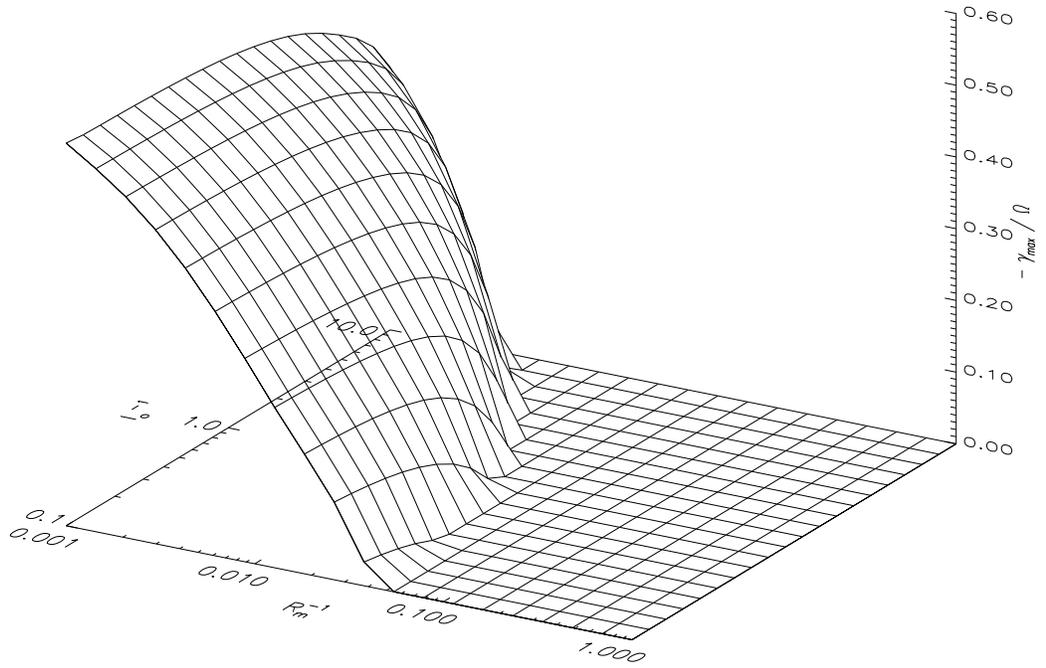}  
   
\caption{Maximum growth rate of the MRI in a stratified disk with
$\beta_c = 10^{2}$ as a function of $l_o^{-1}$ and $R_m$. The 
diffusivity profile is given by equation (\ref{perfil}) with the
surface of the disk placed at $z = H$. }   
\label{fig2}
   \end{figure}

\begin{figure}  
\plotone{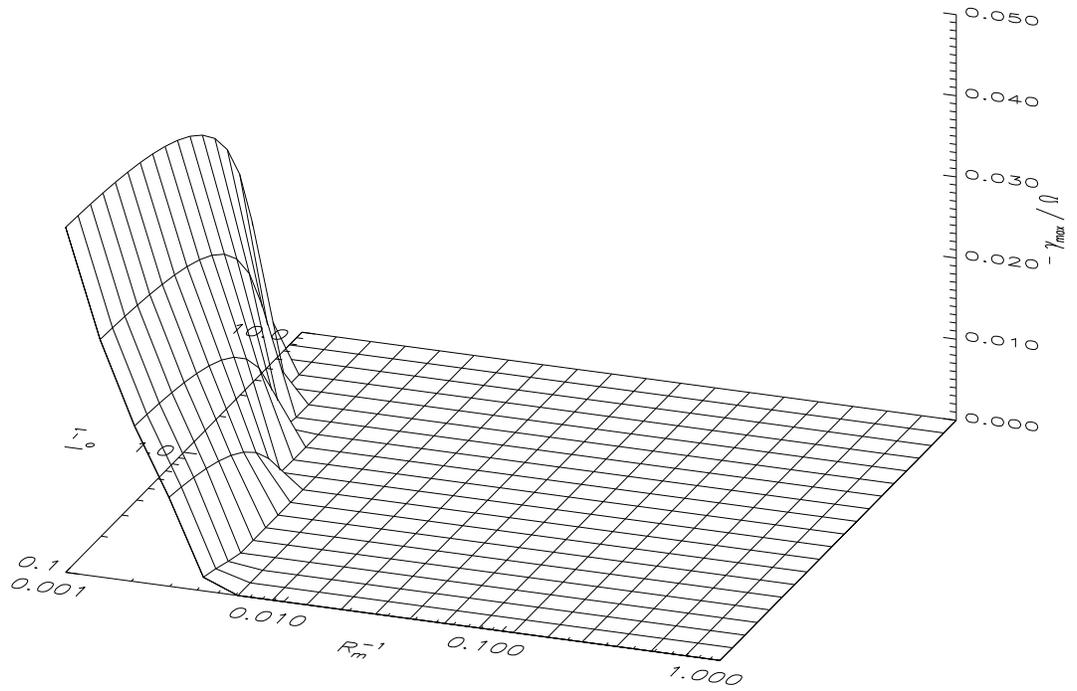}  
  
\caption{Same as figure \ref{fig2} but for an unperturbed magnetic field
corresponding to $\beta_c = 10^{4}$. Notice the change of scale in
the $z$ axis.}  
\label{fig3}    
\end{figure}

Figures \ref{fig2} and \ref{fig3} show the maximum growth rate of the 
MRI in disks with different $\beta_c$, as a function of the
stratification parameters  $R_m$ (a measure of the ohmic magnetic
diffusivity at the surface of the disk) and $l_o$. The range in
both parameters is chosen with the application to PP disks in mind, 
as we discuss in the next section. Figure
\ref{fig2} corresponds to an  unperturbed magnetic field strength with 
$\beta_c = 10^{2}$, which we call the strong magnetic field case.
Figure \ref{fig3} corresponds to $\beta_c = 10^{4}$, referred to as the 
weak magnetic field case. 
In both cases the surface of the disk is 
considered to be at $z=H$.
 
Depending on the strength
of the unperturbed magnetic field 
the effect of the stratification can be determinant for the existence
of the instability. Regions of the disk with characteristic 
diffusive timescales at the surface
less than 10 dynamical timescales, will not be unstable for all 
unperturbed magnetic fields considered, $\beta_c > 10^{2}$. 
The instability will 
also be generally damped if the stratification is strong 
$l_o^{-1} \ga 10$. In the next section we discuss the application
of our results to PP disks.

\begin{figure}  
\plotone{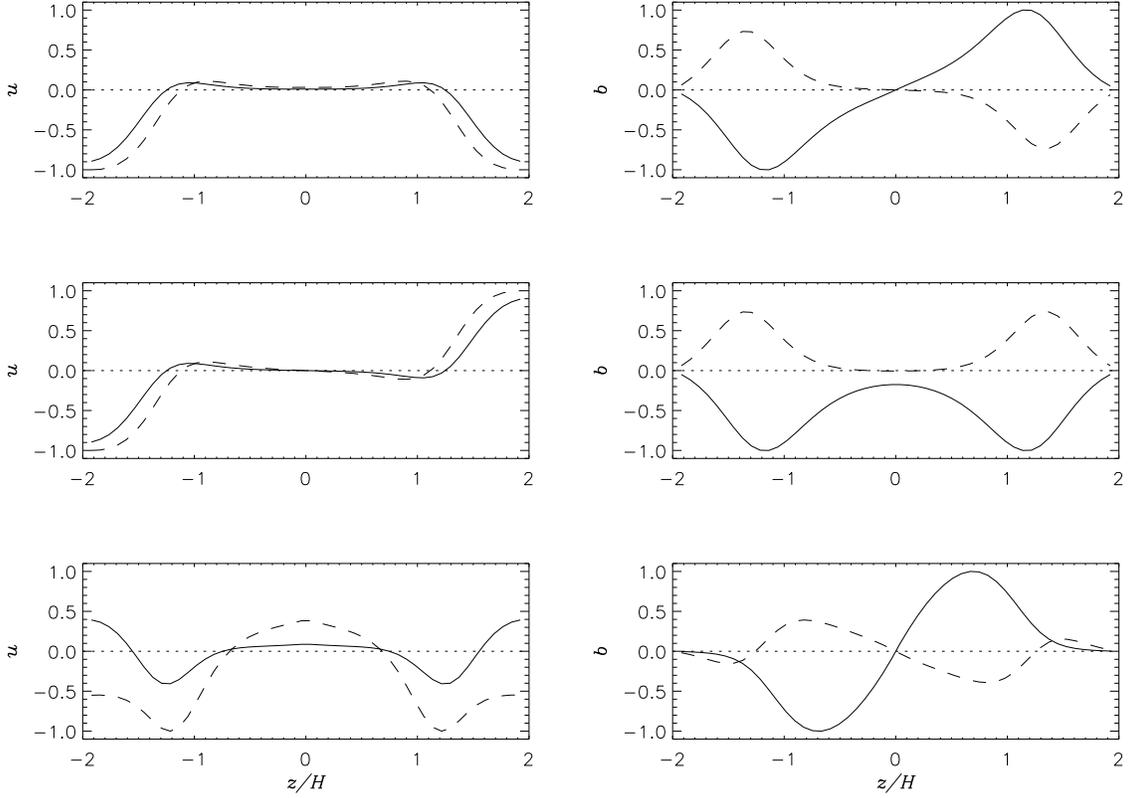}  
  
\caption{Fastest growing modes (from top to bottom) of the MRI in 
a stratified disk with $R_m = 20$, $l_o^{-1} = 1$ and 
$H_t = 2 H$. The unperturbed magnetic field corresponds to
$\beta_c = 10^{2}$. Left and right panels correspond to velocity and 
magnetic field perturbations respectively. In all cases the solid 
(dashed) line indicates the azimuthal (radial) component.
}
\label{fig4}    
\end{figure}

Stratification can also affect the form of the perturbations
for the fastest growing modes. Figure \ref{fig4} shows the 
$z$-dependent amplitude of the three fastest growing unstable 
modes for a region with moderate stratification ($l_o^{-1} = 1$) and
diffusivity near the stabilizing limit ($R_m = 20$). The
surface of the disk is placed at $H_t = 2 H$ and the unperturbed
magnetic field corresponds to $\beta_c = 10^{2}$. An interesting 
feature of this result is the ``stratification'' of the instability, 
i.e. the excitation of the instability preferentially in the 
regions away from the disk midplane. The largest
amplitudes in the linear regime, both for velocity and magnetic 
field perturbations, appear at $z \approx H$.

\section{Discussion}

We have seen that the behavior of the MRI in stratified media depends
crucially on the value and form of the diffusivity profile. It also 
depends on the magnitude of the unperturbed magnetic field originally 
present in the disk. In this section we argue what are the likely
values of this parameters in the context of PP disk models and discuss
plausible consequences of our results.
However, before doing this  
we briefly discuss the implications of modifying some of the 
main assumptions of our analysis.

\subsection{Boundary conditions and field geometry}

As discussed in section 1 the adopted boundary conditions
follow from modeling the exterior medium 
as a hot tenuous corona,
also called a hot halo by Gammie \& Balbus (1994).
The continuity of the stress tensor resulting from 
the perturbations across the disk-halo boundary, and 
the condition that it vanish at infinity lead to 
restrictions (\ref{bcsu}) and (\ref{bcsb}).
While this is apparently a natural choice for isolated
disks, different sets of boundary conditions 
could certainly be constructed and, in principle,
these could affect the results presented so far. 

Gammie \& Balbus (1994) 
propose two other models for the disk exterior: 
a tenuous atmosphere, actually the continuation of the disk 
with an isothermal density profile, and the so-called
rigid conductor model, introduced to mimic the effect of
having a sink of angular momentum outside the disk.
In the former, infinite disk case, no new unstable
modes are found by Gammie \& Balbus (1994) in comparison
to the hot halo model.  
In the second case, when a load is connected to the 
field lines, a new mode is found which can be 
unstable even when all other modes are stable.  
As discussed by Gammie \& Balbus (1994) the new
unstable global mode arises for strong seed magnetic
fields ($\beta_c \sim 1$). Although the results of
Gammie \& Balbus (1994) assume ideal
MHD for the disk material, we believe similar conclusions will be 
found in our case since the introduction of resistivity
has not resulted in the appearance of new unstable modes
(Jin 1996, Sano \& Miyama 1999). 

\begin{figure}  
\plotone{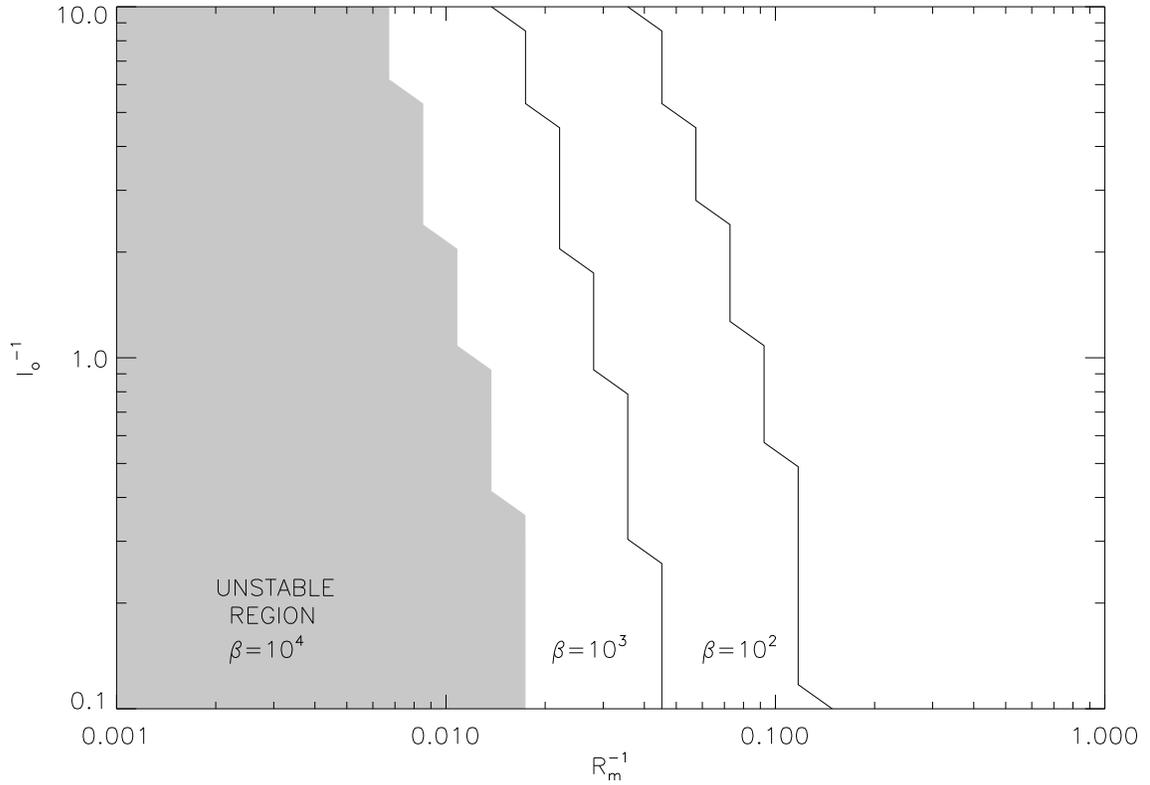}  
 
\caption{Regions of $l_o^{-1} - R_m^{-1}$ parameter
space where the MRI arises for $H_t = 3 H$. The shaded region 
illustrates unstable combinations of the diffusivity 
parameters for $\beta_c = 10^{4}$. The 2 thin solid lines indicate the
boundary of stability in disks with $\beta_c = 10^{3}$ (lower line)
and $\beta_c = 10^{2}$ (upper line).}   
\label{fig5}    
\end{figure}

In addition to the nature of the interface, 
the precise location of the disk-halo boundary is also
assumed. This is done in our model 
through the parameter $H_t$ in equation (\ref{perfil}), 
assumed equal to
the isothermal scale height in the results of the previous section. 
Our choice is motivated by the calculations by 
D'alessio et al. (1999) of the detailed vertical 
structure of PP disks in which the density drops 
rapidly by orders of magnitude after 
$\sim$ 1-2 isothermal scale heights. The termination
of the disk at the scale height
$H$ is also an appropriate choice if one considers 
a polytropic rather 
than isothermal model for the vertical structure. 
Alternatively one could argue that the transition
from the disk to the halo should be located at the 
point where $\beta$ equals unity, i.e. where the magnetic 
field passes from being a weak dynamical agent,
with respect to gas pressure, to being the dominant 
dynamical factor, as is expected in the corona.
Such assumption, with our isothermal disk 
structure, places $H_t$ at approximately 2 and 3 
scaleheights for $\beta_c = 10^{2}$ and $10^{4}$ 
respectively. Figure \ref{fig5} shows the regions 
of stability and instability as a function of 
$R_m$ and $l_o^{-1}$ for different values of 
the parameter $\beta_c$.
In comparison to figures
\ref{fig2} and \ref{fig3}, and to figure \ref{fig7}
to be discussed below, we see that our results 
are not changed drastically, although the trend to
make the disk more unstable as one increases the 
vertical extent is significant. As one incorporates into the 
model regions of lower $\beta$, which are 
more unstable, the damping effect of the poorly 
ionized middle regions is less important.

\subsection{The value of $\beta_c$}

The parameter $\beta_c$ depends on the strength of the unperturbed 
magnetic field seeding the instability. In PP disks a
weak seed magnetic field can be expected as a remnant of the 
disk formation process out of a magnetized molecular cloud.

Measurements of magnetic fields in molecular clouds indicate that the
densest regions, where PP disks are being formed, are commonly threaded
by ordered fields with magnitudes around $\sim 10^{-2}$ Gauss 
(Heiles et al. 1993). If we consider this as the seed field 
for the MRI, our $B$, for the typical midplane
densities $10^{-9}-10^{-11}$ gm/cm$^3$ 
and temperatures $10 - 500$ K predicted for PP disks around a few AU, 
this implies that $10^{2} \la \beta_c \la 10^{5}$. 

Somewhat stronger fields would serve to seed the instability 
near the central star, if it has a significant magnetic field.
However, at radial distances $\ga 1$ AU where the dead zone 
is proposed to exist, such field would most likely be weak,
with $\beta_c$ in the range considered.

\subsection{Ionization state of PP disks}

We now attempt to justify the range of parameters $l_o^{-1}$ and 
$R_m$ explored in section 3 on the basis of the expected 
properties of protoplanetary disks. 

Traditionally the evolution 
of the solar nebula, the archetype of PP disks, has been thought 
of as consisting of 3 stages. In the first  stage, infall from 
the molecular cloud leads to the formation and growth of an 
hydrostatic disk. In this so-called formation stage, efficient 
angular momentum transport resulting from self gravity instabilities 
leads to a fast evolution which, after some $\sim 10^5$ years, 
renders most of the disk stable to gravitational instabilities. 
What follows has usually been called the viscous stage as it is 
believed that the long term evolution of the disk in this phase
is mainly governed by anomalous ``viscous" torques resulting on the 
accretion of mass through the disk and onto the central object.  
Finally, in the dispersal stage the remaining gas and 
dust from the process of planet formation is somehow 
removed. Partly because diffusive, viscous processes act on an ever 
increasing timescale in the disk, it has been suggested that 
external agents dominate this final stage of PP disk evolution
(see Hollenbach et al. 2000 and references therein). 

As mentioned in the introduction, it is in connection to the 
origin of the anomalous torques, dominant through the $\sim 10^7$ 
year lifetime of the viscous stage, that the MRI has become such  
an important part of accretion disk theory. In this paper, 
our aim is to determine whether the MRI will arise in poorly ionized 
portions of PP disks (a prerequisite for generating viscous torques) 
at the beginning and possibly during the viscous stage. Hence, we 
consider models for viscous stage protoplanetary disks.
Over the last 20 years $\alpha$-disk 
models have been constructed addressing the influence of different 
agents on the disk's viscous evolution (see Ruden \& Lin 1986,  Ruden 
\& Pollack 1991, Sterzik \& Morfill 1994, and Reyes-Ruiz \&  
Stepinski 1995, among others). Apart from temporal and detailed 
differences in the structure of the disk, the overall properties 
of the disk generally agree well with the models 
of the Ruden \& Lin (1986). In fact with the exception of the
outermost regions, near the outer boundary of the disk, 
steady state models as those described in detail in 
Stepinski et al. (1993) reproduce satisfactorily the 
properties of PP disks at a given instant. 

In this paper we adopt such steady state models of PP disks
(Stepinski et al. 1993) to describe the radial variation
of physical properties as the disks emerge from the formation 
stage of their evolution. The models incorporate the Shakura \&
Sunyaev (1973) viscosity prescription and temperature dependent 
opacity laws as proposed by Ruden \& Pollack (1991).
The physical properties of the disk are given by a 
sequence of power laws, described in Stepinski et al. (1993),
distributed radially
depending on the two basic parameters of the model:
the turbulence viscosity parameter of Shakura \& Sunyaev,
$\alpha_{ss}$, and the uniform mass accretion rate through the
disk, $\dot{M}$. Values of $\alpha_{ss}$ between 
$10^{-2}$ and $10^{-3}$ are
usually quoted from simulations of the MRI in 
shearing boxes with keplerian differential rotation. As 
such values seem to be also consistent with observations 
of T-Tauri disks (D'alessio et al. 1998) we will present 
results for the limiting values. Again both in models and
observations (see review by Calvet et al. 2000) the mass 
accretion rate ranges from $\sim 10^{-6} M_\odot$/yr, at the
beginning of the viscous stage, to $\sim 10^{-8} M_\odot$/yr
through most of their lifetime.     

A particular choice of model parameters 
determines the profiles of midplane density 
and temperature, surface density, scale height, 
etc. These, and our isothermal vertical structure model,
can be used to calculate the ionization 
degree throughout the disks following 
Stepinski et al. (1993). In the innermost disk
regions, inwards of $\sim 1$ AU, we consider 
thermal ionization of Potassium assumed present
in PP disks with solar abundance. 
As found by Stepinski (1992) in regions with temperature 
above $\sim 10^{3}$ K, the ionization degree is 
high enough to render the disk unstable to the
MRI ($R_m \gg 1$). The development of the MRI
in this region, called the inner active region (IAR) 
by Stepinski (1999) proceeds as analyzed by Gammie 
\& Balbus (1994) and is not addressed in this paper.

We concentrate on the regions outwards of the IAR, 
defined by $r > R_{IAR}$, where $R_{IAR}$ is defined 
as the radius where the temperature drops below 
1000 K. There, ionization is mainly due to the action 
of galactic cosmic rays penetrating through the 
disk surfaces. We obtain an upper limit to the ionization
fraction neglecting the
recombination of electrons onto dust grains and 
considering only ion reactions as a sink of electrons.
Using the ionization and recombinations rates of 
Stepinski (1992), the ionization fraction, $x$, can be 
obtained from the condition of ionization equilibrium,

\begin{equation}
x  \ = \ 4 \times 10^{-18} \rho(z)^{- 1/2} 
         \ {\rm e}^{\frac{-S(z)}{2 S_o}} ,
\end{equation}

\noindent where $S_o$ is the characteristic cosmic ray
attenuation density, $\sim 100$ gm/cm$^2$, and
$S(z)$ is the integrated surface density from the disk
surface to the height $z$:

\begin{equation}
S(z)  \ = \ \int_z^{H_t} \rho_c \ {\rm e}^{- \frac{z^2}{H^2}} \ dz.
\end{equation}

Considering the rapid decrease of the argument, for mathematical
simplicity we approximate this function as:

\begin{equation}
S(\hat{z})  \ \approx \ \frac{\sqrt{\pi}}{2} \ \rho_c \ H \ 
               (1 - {\rm Erf}(\hat{z})),
\end{equation}

\noindent so that the ionization degree can be written as:

\begin{equation}
x(\hat{z})  \ = \ 4 \times 10^{-18} \ \rho_c^{-1/2} \ 
                 h^{-1/2} \  
                {\rm e}^{-(1 - {\rm Erf}(\hat{z}))/l_o} ,
\end{equation}

\noindent where $l_o^{-1} = \sqrt{\pi} \rho_c H/4 S_o$. With 
ohmic diffusivity given by $\eta = 7 \times 10^3 / x$
(Stepinski 1992) we arrive at equation (\ref{perfil}) for
the diffusivity profile.

\begin{figure}  
\plotone{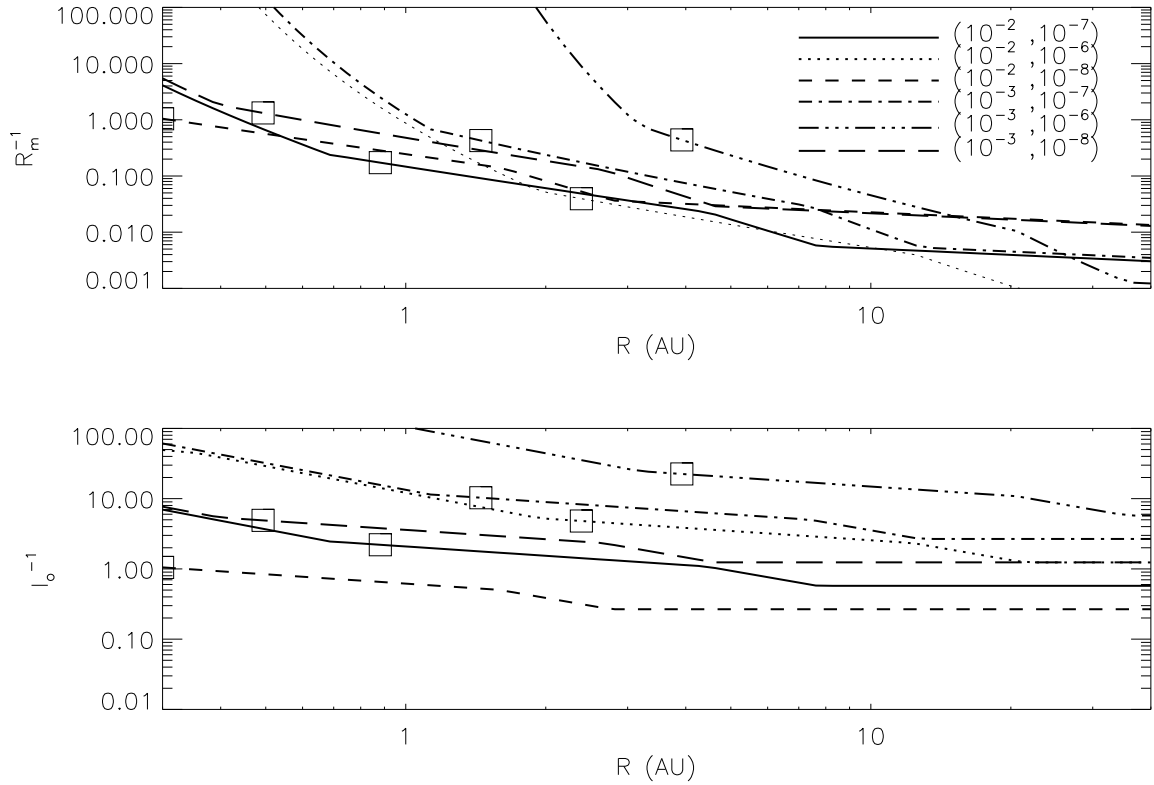}  
 
\caption{Radial profiles of ionization properties, reflected in the parameters
$R_m$ and $l_o^{-1}$ for different models 
of protoplanetary disks. In the legend of linestyles the 
first number indicates the value the $\alpha$ parameter and
the second the mass accretion rate in solar masses per year.
For each model the square indicates the position of the $R_{\rm IAR}$.}   
\label{fig6}    
\end{figure}

\begin{figure}  
\plotone{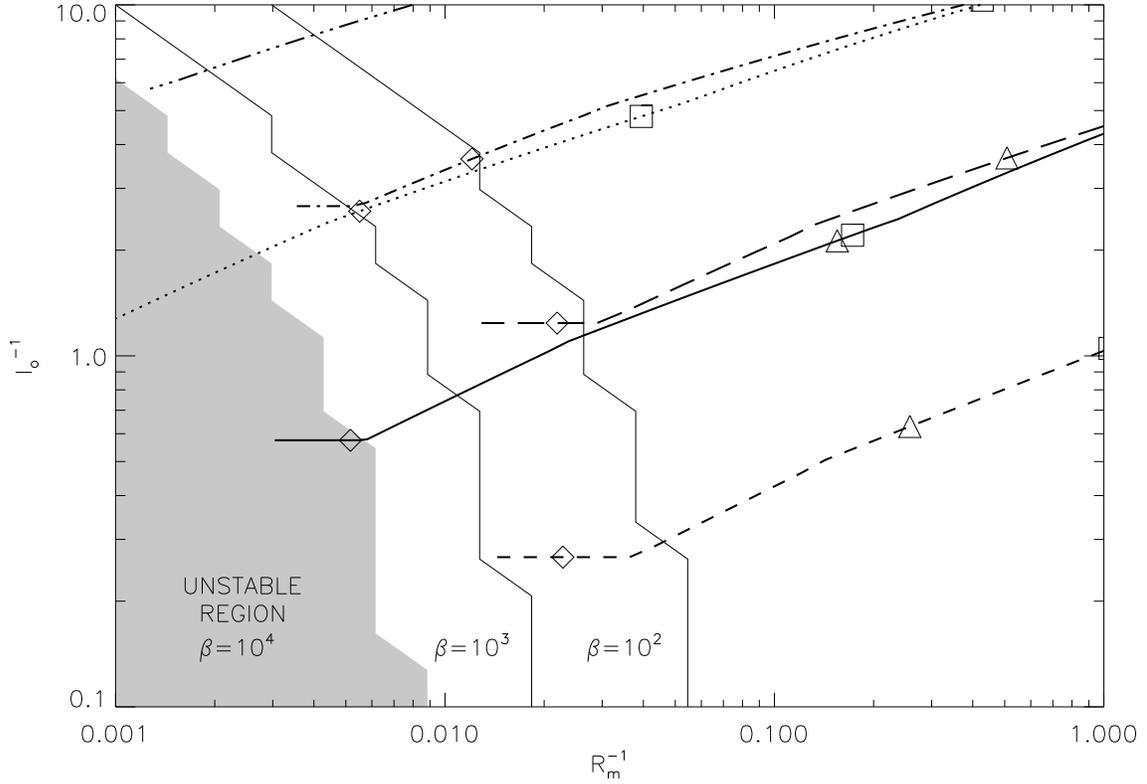}  
  
\caption{Comparison of $R_m$ and $l_o^{-1}$ values for various PP disk 
models with the stability conditions resulting from our analysis. Disk 
models and corresponding linestyles are the same as those presented 
in figure \ref{fig5}. The leftmost edge of such lines 
correspond to the value of $R_m^{-1}$ and $l_o^{-1}$ at 60 AU.
Over each line the square marks the $R_{\rm IAR}$ and the triangle
and diamond mark where $R = 1$ and $R = 10$ AU respectively. 
}
\label{fig7}    
\end{figure}

In figure \ref{fig6} we show the radial profiles of the parameters
$R_m$ and $l_o^{-1}$ characterizing the diffusivity profile, 
for a series of specific models of PP disks. Two values of the 
turbulent viscosity parameter are considered, $\alpha = 10^{-2}$ 
and $10^{-3}$. The mass accretion rate through the disk
takes the values $10^{-6}$, $10^{-7}$ and $10^{-8} \ M_\odot$/yr.
Models with lower $\alpha$ and higher accretion rate are 
characterized by a higher temperature and surface density at
a given radius. For such models the IAR extends farther 
out in the disk in comparison to corresponding models with
higher $\alpha$ and lower $\dot{M}$. Also, in comparison
beyond $R_{\rm IAR}$ their diffusivity is
higher, i.e. smaller $R_m$, as is their stratification reflected 
in the parameter $l_o$. 

We summarize the results of this section in Figure \ref{fig7}
where a direct comparison of the properties of PP disk models and
the regions of stability for different values of $\beta_c$ is shown.
Depending on the value of $\beta_c$, marked radii allow us to 
determine in what regions of the disk the MRI will arise.
For example, for the disk model with $\alpha = 10^{-2}$ and 
$\dot{M} = 10^{-8} M_\odot $/yr (short dashed line), even if the
seed field is as strong as $\beta_c = 10^{2}$, in a broad region
between several AU and the $R_{\rm IAR}$ (square), located at less than 
0.3 AU from Figure \ref{fig6}, the MRI will not arise. For the same model, 
if $\beta_c = 10^{4}$, outwards of $R_{\rm IAR}$ the whole disk will
be stable. A similar analysis 
can be performed for the reader's preferred specific model.

Finally, it is worth stressing that the calculated diffusivities 
represent a lower bound as we neglect recombination onto dust grains, 
which can be dominant if their size distribution is similar to 
that of the ISM (see Reyes-Ruiz \& Stepinski 1995).  

\section[]{Summary and Conclusions}

We have conducted a quasi-global analysis of the 
behavior of the MRI in the linear regime considering 
the effects of diffusivity stratification as that 
expected in protoplanetary disks. As our main result
we find that stratification can add
to the stabilizing effect of ohmic diffusion found in
previous calculations. 

Ohmic diffusivity alone, without stratification, will 
damp the MRI if $R_m \la 10$, regardless of the strength 
of the unperturbed magnetic field. Weak seed magnetic fields,
$\beta_c \ga 10^{4}$, will  become stable if $R_m \la 100$,
i.e. if the diffusion timescale is up to 100 times longer 
than the dynamical timescale. In addition, the stabilizing
effect of stratification is significant when it
is characterized by values of the
parameter $l_o^{-1} \ga 1$, i.e. when the surface density if greater 
than $S_o \approx 100$ gm/cm$^2$.
 
\subsection{Revised model of PP disks}

If the MRI is the only source of 
turbulence and anomalous viscosity in protoplanetary 
disks, our results suggest the necessity of a 
revision of standards models. The calculation 
of properties of these revised models, time dependent
in essence, is out of the scope of this paper. 
However, one can speculate 
on a plausible structure of the revised PP disk models
based on the dynamical agents operating at different radii.

In most of the models, outwards of $R_{\rm IAR}$ there 
is a broad region where the MRI does not operate.
The flow in such region would not be turbulent and,
in absence of anomalous viscous torques, mass would 
accumulate in a stable ``dead'' region (DR). 

In some models,
for example the one with $\alpha = 10^{-2}$ and 
$\dot{M} = 10^{-7} M_\odot$/yr, the instability 
can be excited also in the outermost parts of the
disk. This could occur first in a ``stratified''
manner (as shown in figure \ref{fig4}) due to 
the strong stratification of the diffusivity. 
Whether such excitation leads, in the nonlinear regime,
to a layered accretion scenario as proposed by 
Gammie (1996), is the subject of numerical 
simulations. We propose the existence of a region of
layered accretion (LAR) with 
the cautionary comment that, according to our linear
analysis, significant velocity and magnetic field
perturbations will also be present in the middle 
``dead" zone. These could possibly lead, in the 
nonlinear regime, to somewhat smaller, but 
non-zero viscous torques in the so-called dead zone.
  
\begin{figure}  
\plotone{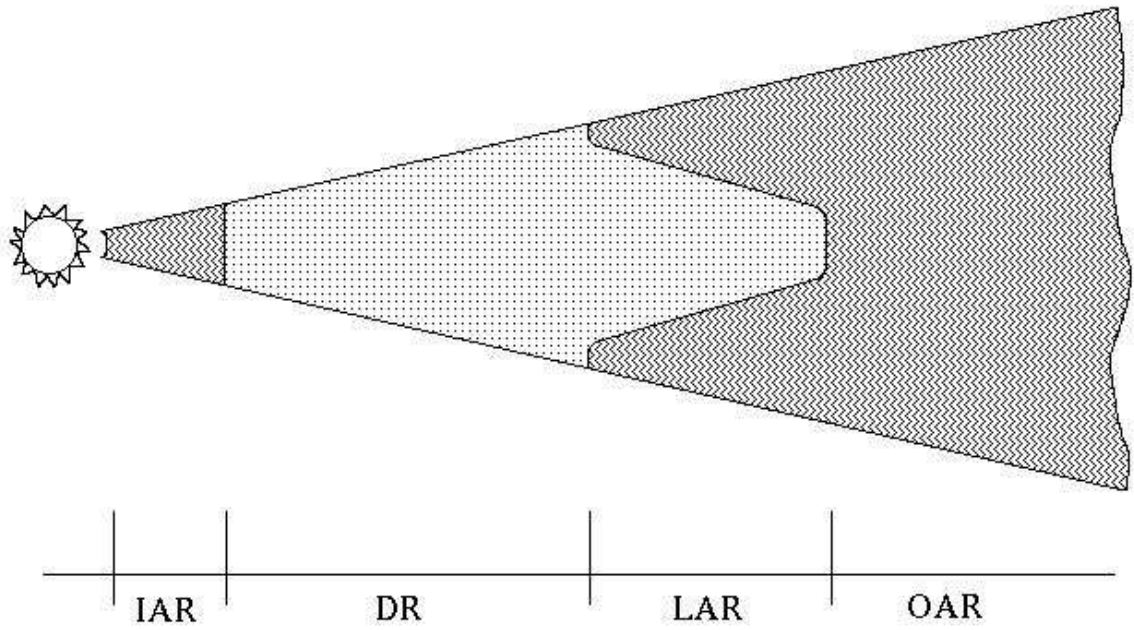}  
 
\caption{Illustration of the general structure of PP disks,
 indicating active and dead regions as suggested by our results.}   
\label{fig8}    
\end{figure}

Regions at a still greater 
distance from the central object are less 
diffusive and less stratified, so the operation of 
the MRI in such regions would be again similar to
that in homogeneous disk and 
turbulence can be generated across the whole vertical
extent of the disk in what we call the outer active region (OAR).
In conclusion, the results derived in this paper with the 
assumptions discussed above lead to a revised model 
for the structure of protoplanetary disks as 
schematically illustrated in figure \ref{fig8}. 

\subsection{Caveats of our analysis}

We end this contribution pointing out potentially 
important inconsistencies in our analysis which
could affect our conclusions.

First, it is clear that our disk models are dynamically
inconsistent. They are derived assuming a constant 
turbulent viscosity parameter $\alpha$ which, as we 
show in this paper, is not justified if turbulence is
due to the MRI. At most, we could argue that such
models reflect the properties of PP disks as they
emerge from the self-gravitating formation stage 
to the viscous evolutionary phase where the MRI is 
believed to be the dominant player in the transport
of angular momentum. Our assumption is made 
as a means to show the necessity of revising the 
standard, constant $\alpha$ models of protoplanetary disks
as well as the layered accretion scenario.

A similar inconsistency is also present in the 
calculation of the vertical diffusivity profile since, 
as shown by Dolginov \& Stepinski (1994), if the 
turbulence in the disk involves a tangled magnetic
field, the shielding of cosmic rays can be significantly 
larger than estimated here. This effect will certainly be 
important in the nonlinear regime of the instability.

%

\vspace{10mm}

\noindent{\bf Acknowledgements}  
The author is thankful to Adriana Gazol for helpful comments
at the beginning of this work.
This work has been supported by CONACYT project J22990E.

\clearpage
 %
%
%
%

\clearpage
 
%
%
%
%

\clearpage

%
%
%
%
 
\clearpage

%
%
%
%

\clearpage

%
%
%
%
\clearpage  

%
%
%
%

\clearpage

%
%
%
%

\clearpage

%
%
%
%
 
\end{document}